\documentclass[twocolumn]{aastex631}
\usepackage[T1]{fontenc}
\usepackage{xcolor}        
\usepackage{tablefootnote}
\usepackage{amsmath}
\usepackage{comment}
\usepackage{color}

\begin{document}

\title{Image Morphology of Misaligned Magnetically Arrested Disks}

\author[0009-0006-4435-0418]{Samuel Feyan}
\affiliation{Center for Astrophysics $|$ Harvard \& Smithsonian, 60 Garden Street, Cambridge MA 02138, USA}
\affiliation{Black Hole Initiative at Harvard University, 20 Garden Street, Cambridge, MA 02138, USA}
\author[0000-0002-0393-7734]{Cora Prather}
\affiliation{Center for Astrophysics $|$ Harvard \& Smithsonian, 60 Garden Street, Cambridge MA 02138, USA}
\affiliation{Black Hole Initiative at Harvard University, 20 Garden Street, Cambridge, MA 02138, USA}
\author[0000-0001-9939-5257]{Dominic O. Chang}
\affiliation{Center for Astrophysics $|$ Harvard \& Smithsonian, 60 Garden Street, Cambridge MA 02138, USA}
\affiliation{Black Hole Initiative at Harvard University, 20 Garden Street, Cambridge, MA 02138, USA}
\author[0000-0002-7179-3816]{Daniel C.\ M.\ Palumbo}
\affiliation{Center for Astrophysics $|$ Harvard \& Smithsonian, 60 Garden Street, Cambridge MA 02138, USA}
\affiliation{Black Hole Initiative at Harvard University, 20 Garden Street, Cambridge, MA 02138, USA}
\author[0000-0002-2825-3590]{Koushik Chatterjee}
\affiliation{Canadian Institute of Theoretical Astrophysics, University of Toronto, 60 St. George Street, \\ Toronto, ON M5S 3H8, Canada}
\author[0000-0002-4120-3029]{Michael D.\ Johnson}
\affiliation{Center for Astrophysics $|$ Harvard \& Smithsonian, 60 Garden Street, Cambridge MA 02138, USA}
\affiliation{Black Hole Initiative at Harvard University, 20 Garden Street, Cambridge, MA 02138, USA}

\begin{abstract}
Horizon scale images probe the plasma dynamics and spacetime in the vicinity of accreting black holes. Interpreting these images requires models that connect image morphology to the structure of the underlying accretion flow. Disk misalignment is expected in many systems, yet most horizon scale image analyses assume that the emitting flow is aligned with the black hole spin. 
We investigate whether tilted accretion flows produce image features distinguishable from axisymmetric flows and whether their effects can be absorbed by changes in aligned model parameters. 
Using general relativistic magnetohydrodynamic simulations of tilted magnetically arrested disks spanning black hole spins $a^*=-0.9375$ to $0.9375$ and initial disk tilts $0^\circ$ to $30^\circ$, we generate synthetic images and quantify their morphology through fits to an aligned, semi-analytic emission model. 
Using this approach, we show that high spin prograde models show little sensitivity to disk tilt, as the inner emitting flow aligns with the black hole spin. 
In contrast, retrograde tilted disks develop warped, non-axisymmetric inner structure that distorts the inner shadow and strongly alters image morphology.
While the aligned semi-analytic model reproduces most prograde images well, retrograde tilted disk images are fit less accurately and yield biased inferred parameters, especially for spin. The image morphologies indicate that tilt affects horizon scale images only when misalignment survives into the near horizon emitting region, motivating image models that include disk tilt as an explicit degree of freedom when alignment is uncertain.
\end{abstract}


\section{Introduction}

Accretion disks around black holes are often modeled under the assumption that the disk angular momentum is aligned with the black hole spin axis. Such alignment, however, may not be generic. In the context of supermassive black holes (SMBHs), misalignment could be a natural outcome of their evolution. Tidal disruption events can supply material to the SMBH at random orientations relative to the spin axis, while SMBH mergers can reorient the spin of the remnant relative to gas supplied on larger scales by the host galaxy \citep{2024arXiv240410052F}. Although tidal interactions eventually tend to align a tilted disk \citep{1978Natur.275..516R}, for TDEs these alignment timescales can be comparable to the system's observational lifetime \citep{1996MNRAS.282..291S, 2016MNRAS.455.1946F}, and for SMBHs, repeated merger events can generate recurrent tilted disk configurations \citep{2000ApJ...537..152K}.

Observations of both X-ray binaries and active galactic nuclei support the possibility of disk misalignment. In both systems, observations suggest that the large scale accretion disk may not always be aligned with the black hole spin axis, assuming that the misaligned jet traces the spin direction \citep{1997ApJ...477..876O,2005ApJ...618..618K}. In some AGN, reflection spectroscopy further reveals warped disk geometries consistent with misaligned accretion flows \citep{2016MNRAS.457.1568M}. Additionally, some quasi-periodic oscillations (QPOs) have been interpreted with models that invoke disk spin misalignment \citep{2009MNRAS.397L.101I}.

Misaligned accretion disks exhibit a broad range of phenomena not present in their aligned counterparts. Misalignment induces nodal precession of particle orbits through the Lense-Thirring effect \citep{1918PhyZ...19..156L}, with a characteristic rate that scales steeply with radius and linearly with spin, $\Omega \propto a^*/r^3$. As a result, a tilted disk composed of noninteracting rings would undergo differential precession, twisting and warping the flow \citep{2012ApJ...757L..24N}.

Real disks, however, couple neighboring annuli through internal stresses. This response depends on disk thickness, where thin disks can undergo Bardeen-Petterson alignment, in which the innermost disk aligns with the black hole spin while the outer disk remains misaligned \citep{1975ApJ...195L..65B}. Thick disks, in contrast, can support standing bending wave patterns in which the misalignment varies with radius \citep{1997MNRAS.285..394I}. GRMHD simulations of thick tilted disks also show high density plunging streams and standing shocks near the black hole \citep{2007ApJ...668..417F}. The associated shock heating can raise the inflow temperature and dominate the resulting emission signatures \citep{2013MNRAS.432.2252D}.

Magnetic field structure can significantly alter the dynamics of tilted disks. In magnetically arrested disks (MADs) \citep{2003ApJ...592.1042I,2003PASJ...55L..69N}, magnetic flux accumulation can power jets that torque the inner accretion flow and align to the black hole spin axis, especially for rapidly spinning black holes \citep{2018MNRAS.474L..81L, 2026arXiv260527334S}. Additionally, some MAD simulations exhibit little or no disk precession, as the disk is dominated by magnetic stresses \citep{2025PhRvD.112f3013C}. The MAD alignment behavior contrasts with tilted standard and normal evolution (SANE) disks \citep{2012MNRAS.426.3241N,2013MNRAS.436.3856S}, which generally show global precession and do not show strong alignment near the black hole \citep{2019ApJ...878...51W, 2026arXiv260527334S}.

The Event Horizon Telescope (EHT) can now directly image the innermost regions of these accretion flows \citep{2019ApJ...875L...1E}. However, disk misalignment was not included, or very limited, in the original GRMHD modeling of Sgr A* and M87* \citep{2019ApJ...875L...5E,2022ApJ...930L..16E}. In \citet{2022ApJ...930L..16E}, only two tilted disks are studied, with $a^*=0.94$ and initialized in a SANE state. They find a warped and non-axisymmetric disk and jet due to Lense-Thirring precession alongside increased variability and an increase in the 2.2 $\mu$m flux density with tilt.  These trends are consistent with the findings of \citep{2022ApJ...926..136W}, which study a similar $a^*=0.9$ SANE tilted disk. Given the distinct fluid structure associated with tilted disks, it is important to determine whether misalignment can produce different image morphologies that imprint on EHT observables.

\subsection{Relation To Previous Tilted Disk Imaging Studies}

Previous tilted disk imaging studies have shown that misalignment can imprint on near horizon emission. Synthetic images of tilted Sgr A* models have demonstrated that non-axisymmetric standing shocks can produce a wide range of image morphologies and introduce a dependence on observer azimuth \citep{2013MNRAS.432.2252D}, {as the disk is no longer azimuthally symmetric}. Additionally, high spin tilted SANE disks can change the apparent size, shape, and eccentricity of the central brightness depression, and bias physical parameters inferred from aligned models \citep{2020ApJ...894...14W}.

In the context of M87*, \citet{2020MNRAS.499..362C} ray traced tilted disk simulations at 43 and 230 GHz and showed that {warping of the} disk and jet {from tilt} can leave observable signatures on horizon scale images. They further argued that misaligned disk geometries may be relevant for M87* and found that tilted systems can exhibit significant 230 GHz flux variability. However, these simulations did not reach a fully saturated MAD state and {their disks never reached alignment}, so they did not {fully investigate} whether magnetic torques in magnetically arrested disks suppress tilted image signatures through inner disk alignment.

Thus, existing tilted disk image studies have primarily {consisted of} SANE or not fully MAD flows, leaving fully magnetically arrested tilted disks comparatively unexplored. Studies of tilted MADs have focused primarily on accretion dynamics, magnetic torques, jet behavior, and magneto-spin alignment rather than image morphology \citep{2026arXiv260116370G,2025PhRvD.112f3013C}. Existing tilted disk image studies have therefore established that tilt can imprint on image morphology, but it remains unclear whether these signatures persist in MAD disks, where magnetic torques can realign the inner emitting flow in some regions of the spin tilt parameter space.

{The present work} extends previous studies by carrying out a survey of tilted MAD image morphology across black hole spin and disk tilt. Rather than focusing on a single tilted configuration, we compare a grid of MAD simulations spanning both prograde and retrograde black hole spins, with initial disk tilts ranging from $0^\circ$ to $30^\circ$. The simulation grid allows us to assess how tilt imprints on image signatures and whether those signatures depend on spin direction and magnitude. {We also evaluate} whether tilted MAD images can be mistaken for aligned accretion flows. In particular, we use {ray traced} images and emissivity maps to connect image morphology to the underlying {accretion flow}, and we compare the resulting images with semi-analytic aligned models to assess possible degeneracies {and parameter bias when analyzed under the assumption of an aligned accretion flow}.

\subsection{Outline of Paper}

The remainder of this paper is organized as follows. In Section \ref{sec:methods}, we describe the numerical methods used in this work. We introduce the tilted MAD GRMHD simulation suite in Subsection \ref{sec:grmhd_simulations}, describe the ray tracing procedure and viewing geometry in Subsection \ref{sec:ray_tracing}, and present the aligned semi-analytic model and image fitting procedure in Subsection \ref{sec:semi_analytic_model}. In Section \ref{sec:results}, we examine how image morphology varies with black hole spin and disk tilt, and relate these trends to the underlying fluid and emission structure. {Additionally, we} assess how well the tilted images can be reproduced by an aligned semi-analytic model. We summarize our main conclusions and discuss their implications for horizon scale image modeling in Section \ref{sec:conclusion}.

\section{Methods}

\label{sec:methods}

\subsection{GRMHD Simulation Suite}
\label{sec:grmhd_simulations}

This work uses a suite of tilted disk simulations produced with KHARMA (Kokkos-based High Accuracy Relativistic Magnetohydrodynamics with Adaptive Mesh Refinement) \citep{2024arXiv240801361P} based on the HARM scheme \citep{2003ApJ...589..444G}. Throughout this work, we use gravitational units, with $r_g = GM/c^2$ and $t_g = GM/c^3$. The simulations have a resolution of $N_r \times N_\theta \times N_\phi = 384 \times 192 \times 192$ in modified spherical Kerr-Schild coordinates. The computational domain extends from 5 zones inside the event horizon to an outer radius of $1000\,r_g$ and spans the full $2\pi$ azimuthal domain. We use outflow radial boundary conditions, reflecting polar boundary conditions, and periodic boundary conditions in azimuth. 
{Reflecting polar boundary conditions can lead to artificial shearing or torquing of the fluid through the coordinate pole in the most tilted disks, where the disk and pole can intersect.  This boundary condition was chosen due to limitations in the boundary conditions implemented in KHARMA when the simulations were run.  }
The black hole spin axis is aligned with the coordinate polar axis. Additionally, the computational grid has a Lorentz factor ceiling of 5.

\begin{figure*}
    \centering
    \includegraphics[width=\textwidth]{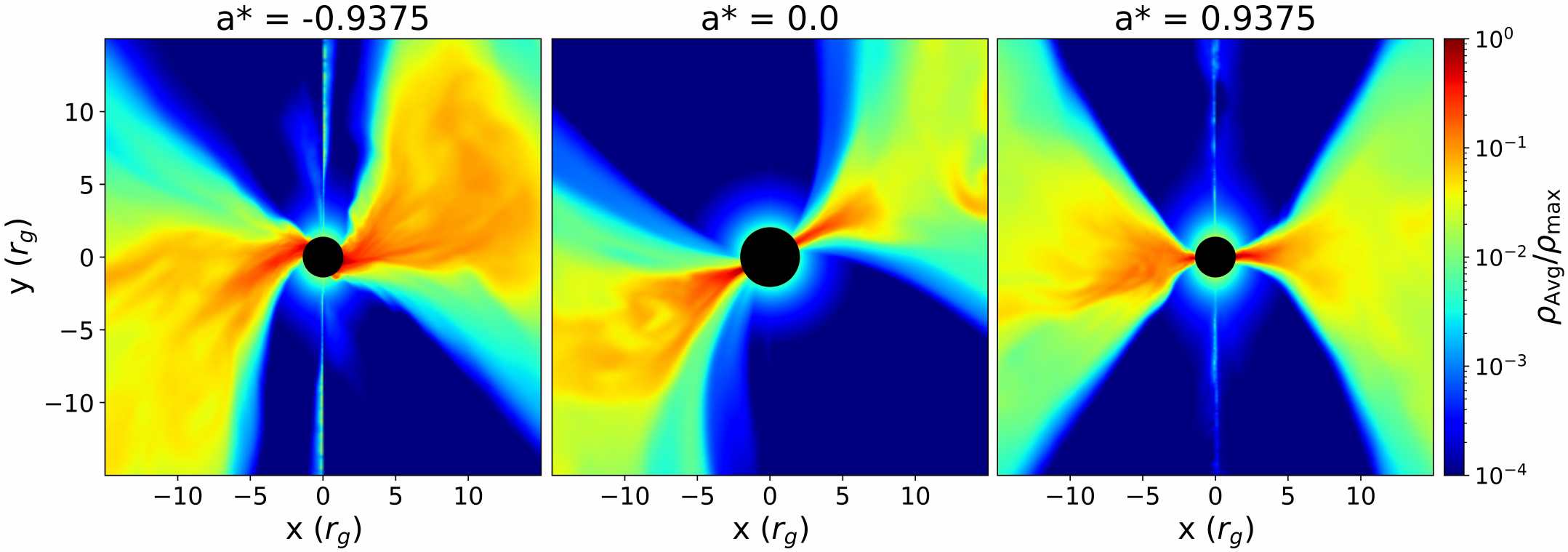}
    \caption{Grid of the fixed two dimensional cross section of the average fluid density, $\rho_{\text{Avg}}$, normalized to the maximum fluid density, $\rho_{\text{max}}$, of the tilt 30 simulations, for $a^*=-0.9375, 0.0,$ and $0.9375$. The fluid average is taken over 100 $t_g$ from 10,000 $t_g$ to 10,100 $t_g$, over 20 snapshots. We use a shorter averaging interval than the 1,000 $t_g$ imaging interval to suppress short timescale fluctuations while preserving coherent structure in the accretion flow. A longer average would obscure these features, particularly in the more variable retrograde simulations, even though the disks show no long term secular evolution in the innermost region at this stage of the simulation, {as shown in \autoref{DiskProfile_t00_t10_t20_t30_r5_to_15rg_theta_vs_tg_all_spins}}. The $a^*=-0.9375$ panel has a persistently warped accretion flow very near the horizon. This is in contrast to the $a^*=0.0$ panel which remains completely misaligned and the $a^*=0.9375$ panel which has a very nearly aligned accretion flow.  
    }
    \label{rho_tilt30_spin-0.9375_2000_2020_avg}
\end{figure*}

The simulations are initialized with a standard Fishbone-Moncrief torus \citep{1976ApJ...207..962F}, with an inner edge at 15 $r_g$, a pressure maximum at 29.5 $r_g$, and an adiabatic index $\Gamma = 5/3$.  We consider a grid of 20 simulations, spanning black hole spins of $a^* = -0.9375$, $-0.5$, $0.0$, $0.5$, and $0.9375$ and initial disk tilts of $0^\circ$, $10^\circ$, $20^\circ$, and $30^\circ$, where negative spin values denote retrograde accretion flows.  The initial disk tilt is defined as the angle between the black hole spin axis and the initial torus angular momentum vector.  For the tilted cases, we rotate the initial torus angular momentum vector relative to the black hole spin axis while leaving the coordinate system fixed.  The magnetic field is initialized with a large scale poloidal magnetic field loop and is rotated with our torus, with vector potential

\begin{equation}
\begin{aligned}
    A_{\phi} =
    \max\Bigg[
        &\frac{\rho}{\rho_{\max}}
        \left(\frac{r}{r_0}\sin\theta\right)^3 \\
        &\times e^{-r/400}
        - 0.2,\,
        0
    \Bigg],
\end{aligned}
\end{equation}
where $\rho$ is the gas density, $\rho_{\text{max}}$ is the maximum initial plasma density, and $r_0$ is the inner edge of the accretion disk \citep{2022ApJS..259...64W}.

All simulations {are} evolved to 11,000 $t_g$ and we generate synthetic images from the final 1,000 $t_g$, corresponding to 10,000 - 11,000 $t_g$.  The simulation and imaging intervals are chosen to sample the inner accretion flow while maintaining a computationally feasible survey across the full spin and tilt parameter space.  We ray trace 200 snapshots from each simulation in intervals of 5 $t_g$ and time average the resulting images.

We verify that each simulation reaches a magnetically arrested state before the imaging interval by measuring the dimensionless magnetic flux threading the black hole event horizon.  During the imaging interval, the measured magnetic fluxes are roughly consistent with the saturated MAD values reported by \citep{2022MNRAS.511.3795N}.  The mass accretion rate onto the black hole also reaches a quasi-steady state before the imaging period.  Together, the magnetic flux and accretion rate diagnostics indicate that the selected interval samples the {relaxed} inner accretion flow rather than the initial configuration.

Although the outer disk is not expected to reach an {equilibrium} state over the full domain, the horizon scale image morphology is dominated by emission in the innermost ${\sim}$10 $r_g$ \citep{2019ApJ...875L...5E}, where the dynamical timescales are significantly shorter. The emissivity maps in \autoref{emissivity_origin_phi_slice_grid_230ghz_rhigh20_rlow1} and \autoref{grid_emiss_230ghz_rh20_rl1_end02200} support this interpretation by showing that the relevant emission originates very near the black hole. 
{At 10 $r_g$, our simulations are run for ${\sim} 55$ Keplerian orbits and the disk warp may not have had enough time to fully settle.}
{To address this}, we examine the time evolution of the disk tilt angle across the inner emitting region in Appendix \ref{app:disk_tilt_analysis}. The tilt in our simulations remains predominantly in a quasi-steady state {by the time we reach} the imaging interval, indicating that the inner disks are no longer undergoing substantial alignment {over the imaging period}.

\subsection{Ray Tracing}
\label{sec:ray_tracing}

To post process fluid snapshots from the GRMHD simulations into synthetic images, we use the general relativistic radiative transfer code \texttt{ipole} \citep{2018MNRAS.475...43M}. Each snapshot is ray traced independently in the fast light approximation, including synchrotron emission and absorption. We generate total intensity images at an observing frequency of 230 GHz. All images are generated on a 320 $\times$ 320 pixel image plane with a field of view of 80 $\times$ 80 $\mu$as.

The GRMHD simulations evolve the plasma as a single fluid and do not separately track the ion and electron temperatures. Because the accretion flow is assumed to be radiatively inefficient, radiative losses are negligible and are not included during the simulation. We assume a quasi-neutral, fully ionized plasma, such that the proton and electron number densities are approximately equal. Electrons radiate synchrotron emission much more efficiently than ions because of their substantially smaller mass and therefore dominate the observed 230 GHz emission. Thus, we assign the electron temperature in post processing using the ion-to-electron temperature prescription of \citep{2019ApJ...875L...5E}.

\begin{equation}
    \frac{T_i}{T_e} = \frac{R_{low} + R_{high} \beta^2}{1 + \beta^2}
\end{equation}

Here $\beta = p_{\rm gas}/p_{\rm B}$ is the plasma beta, where $p_{gas}$ is the gas pressure, $p_B = b^2/2$ is the magnetic pressure, $b^2 = b^\mu b_\mu$, and $b^\mu$ is the magnetic field four vector measured in the fluid frame. 
$T_i$ and $T_e$ are the ion and electron temperatures respectively. We consider one electron temperature prescription, $R_{\rm low}=1$ \& $R_{\rm high}=20$, as our fiducial model.

The ray tracing is scaled to M87* by adopting a black hole mass of $6.2 \times 10^9$ M$_{\odot}$ \citep{2011ApJ...729..119G} and distance of $16.9 \times 10^6$ pc \citep{2019ApJ...875L...5E}. For each simulation, {we normalize the flux} using the time-averaged 230\,GHz images to match the total flux of M87* at $\sim$ 0.5 Jy.

We set the observer inclination to $163^\circ$ for the prograde and non-spinning simulations and to $17^\circ$ for the retrograde simulations, where this inclination angle is the polar angle measured from the black hole spin axis. This choice ensures that the projected fluid velocities will have the same orientation on the sky across spin. The inclination comes from the large scale observed jet position angle {of M87*} \citep{2016A&A...595A..54M}, although \citet{2020MNRAS.499..362C} showed that the jet doesn't necessarily follow the black hole spin axis at large distances.

\begin{figure*}
    \centering
    \includegraphics[width=\textwidth]{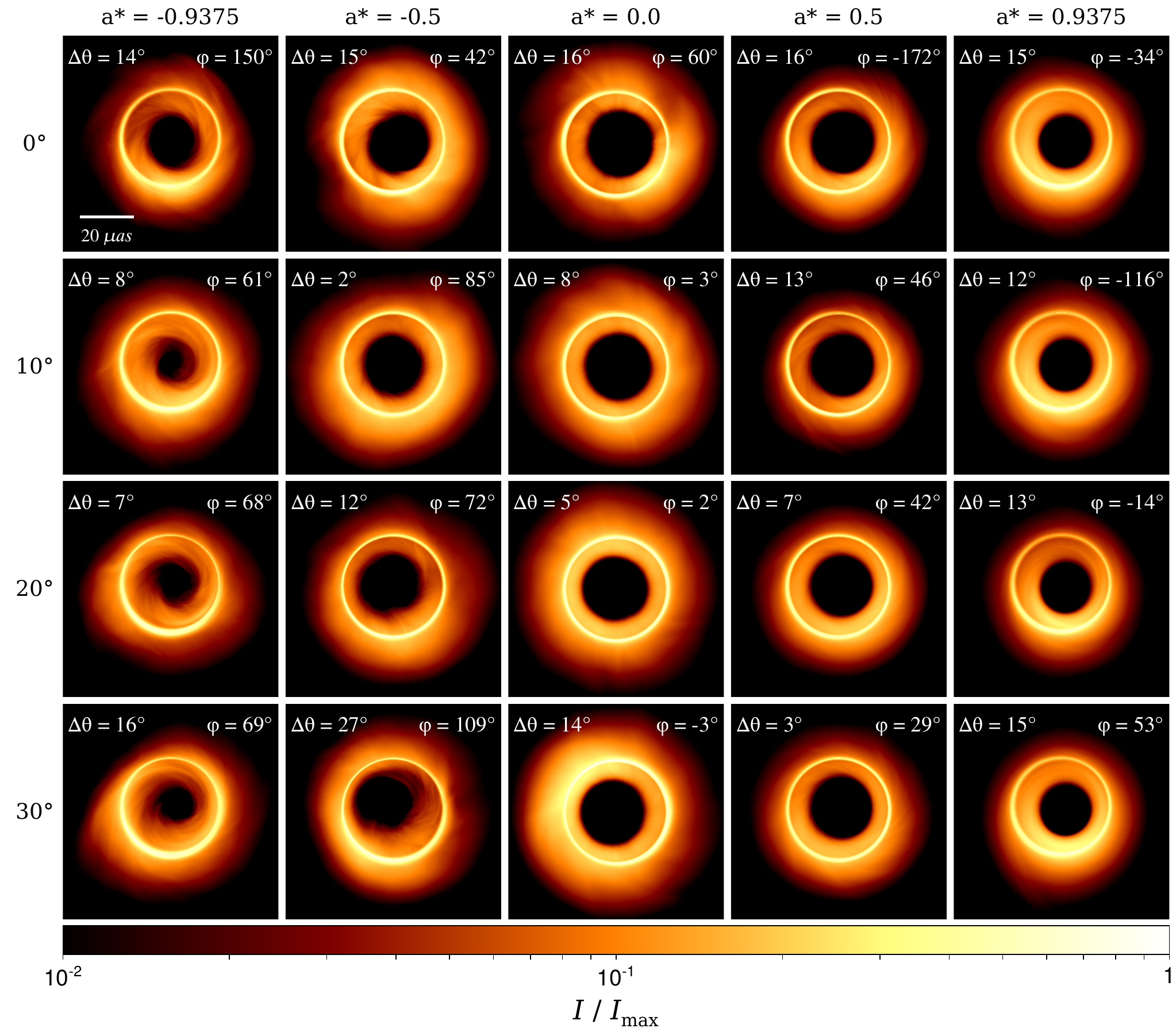}
    \caption{Time averaged 230 GHz images of twenty MAD GRMHD simulations with varying spin ($a^* = -0.9375$ to $a^* = +0.9375$ from left to right) and tilt angle ($0^\circ$ to $30^\circ$) from top to bottom. Each image is at $R_{\rm high} = 20$ and $R_{\rm low}=1$ and the black hole spin vector projected on the plane of the sky is aligned to the east-west direction. The field of view for each image is 80 $\mu$as. {In the upper left of each panel, $\Delta \theta$ is defined as the difference between the disk normal averaged over our imaging period and our viewing angle $|17^\circ - \langle \theta_{\rm disk} \rangle|$. In the upper right, $\varphi$ is the azimuthal angle used in the ray tracing procedure described in Subsection \ref{sec:ray_tracing}. }
    }
    \label{log_grid_230ghz_rh20_rl1_end02200}
\end{figure*}

{It is important to note that this choice in inclination guarantees that we are oriented consistently across simulations relative to the spacetime. Changing the observer position with respect to the spacetime can imprint on the photon ring and shadow position in the image \citep[e.g.,][]{2004A&A...423..787T,2010ApJ...718..446J}. We however are not oriented consistently relative to the disk normal or the jet, as our tilted disks have not necessarily fully aligned to the black hole spin axis. We quantify this difference in the inclination by finding the difference between the normal of the disk calculated in \autoref{app:disk_tilt_analysis} and our choice in inclination and list them in \autoref{log_grid_230ghz_rh20_rl1_end02200}.}

As the tilted simulations are no longer axisymmetric, the observer azimuthal angle must also be specified. We choose this angle using the total rotation angle of the inner jet around the polar axis, known as the precession angle. Following the prescription of \citet{2018MNRAS.474L..81L}, we compute the magnetic pressure weighted centroid of the upper jet in Cartesian coordinates. We identify the upper jet using the magnetization cut $\sigma>1$ and the coordinate polar angle cut $0<\theta<\pi/2$, where $\sigma=b^2/\rho c^2$.

Although the simulations are evolved in modified spherical Kerr-Schild coordinates, we compute the jet centroid using Cartesian spatial coordinates $x^i\in\{x,y,z\}$ to avoid coordinate ambiguities near the polar axis. For each radial shell $\Delta r$, the centroid coordinate is computed as
\begin{equation}
    x^i_{\rm jet}(r,t)
    =
    \frac{
        \int_{\Delta r}
        p_B x^i \sqrt{-g}\,d\theta d\phi
    }
    {
        \int_{\Delta r}
        p_B \sqrt{-g}\,d\theta d\phi
    },
    \label{eq:jet_centroid}
\end{equation}
where the integral is restricted to cells satisfying the upper jet selection.

The jet precession angle $\varphi$ is then defined as the azimuthal angle of this centroid in the Cartesian basis. {This $\varphi$ is defined such that the disk normal lies along $\varphi = 0, 180$ depending on whether the observer is viewing the upper side or lower side of the disk. This angle can also be interpreted as the integrated precession angle from the start of the simulation.}
For the ray traced images, we use the time averaged value of $\varphi$ during the imaging period over the inner jet region, $2~r_g~{\leq}~r~{\leq}~5~r_g$, to set the observer azimuthal angle. This choice orients the observer azimuthal position consistently with the jet base across the simulation suite and accounts for jet precession in the tilted models. As the prograde and non spinning simulations have observer inclinations of $163^\circ$ and are viewing the lower jet, we offset this upper jet precession angle by $180^\circ$ accordingly.  {These azimuthal angles calculated for each simulation are listed in \autoref{log_grid_230ghz_rh20_rl1_end02200}.  }

We exclude emission from the jet region by applying the same magnetization cut, $\sigma>1$. The magnetization cut removes emission from the mass loaded jet that could otherwise contaminate the image. We also excise a narrow region around the poles from $\theta < 3^\circ$ and $\theta > 177^\circ$ to remove artifacts from the polar boundary conditions that may otherwise not be fully removed by the $\sigma$ cut in high tilt images. Additionally, we rotate all images so the projected black hole spin axis points along the east-west direction in the image plane.

\begin{figure*}
    \centering
    \includegraphics[width=\textwidth]{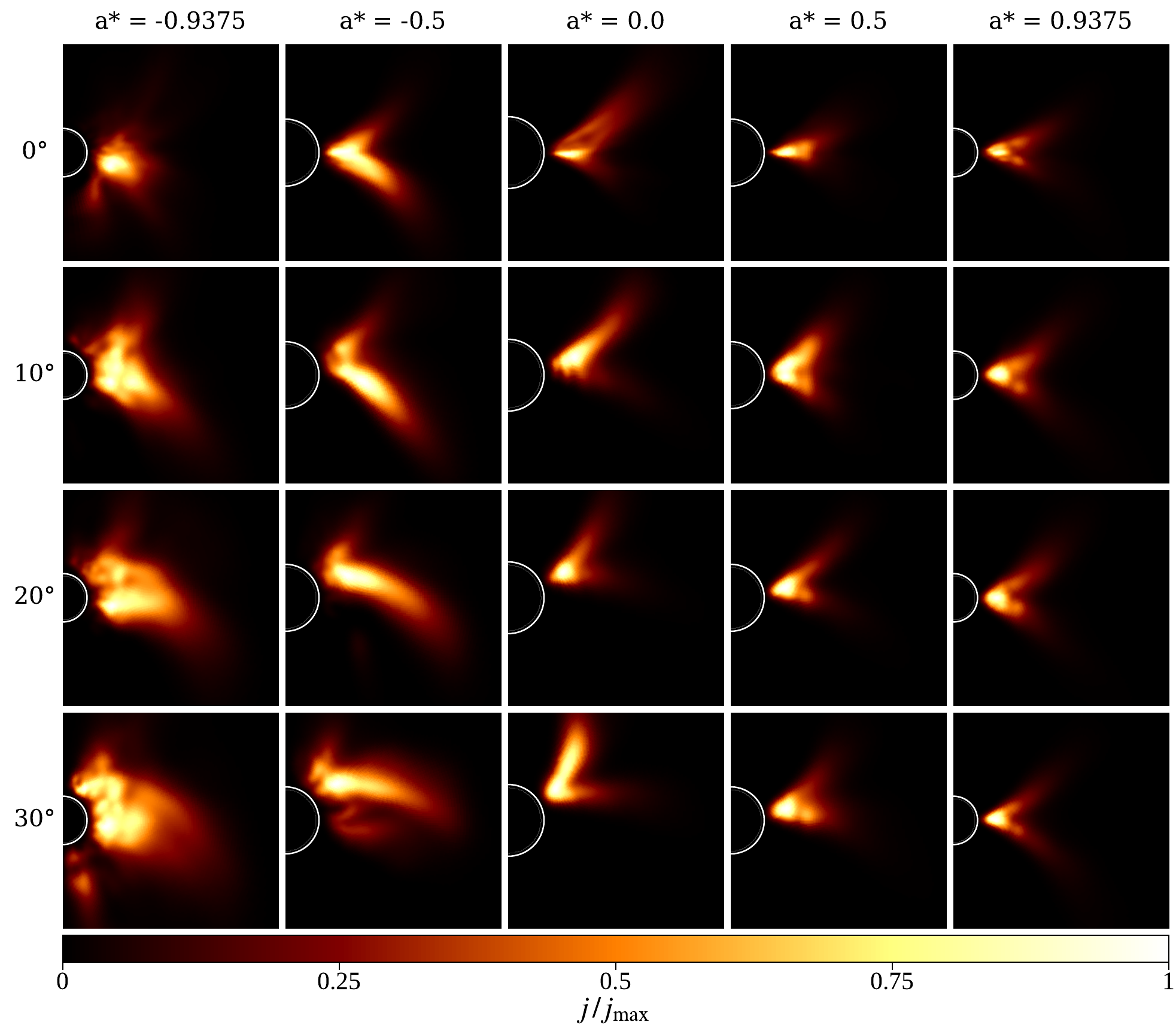}
    \caption{Binned location of the point of origin, or emissivities, for the photons that make up the individual images from \autoref{log_grid_230ghz_rh20_rl1_end02200} at 230 GHz with $R_{\rm high} = 20$ and $R_{\rm low} = 1$. The image field of view is 12 $r_g$ across on both the vertical and horizontal axis.  The emissivities are averaged in time and taken at a particular azimuthal slice, corresponding to $\varphi$ to maximize the apparent tilt. In each panel, the black hole spin axis is aligned to the vertical figure axis. Plotted in white over the image is the outer event horizon of the black holes.
    }
    \label{emissivity_origin_phi_slice_grid_230ghz_rhigh20_rlow1}
\end{figure*}

\begin{figure*}
    \centering
    \includegraphics[width=\textwidth]{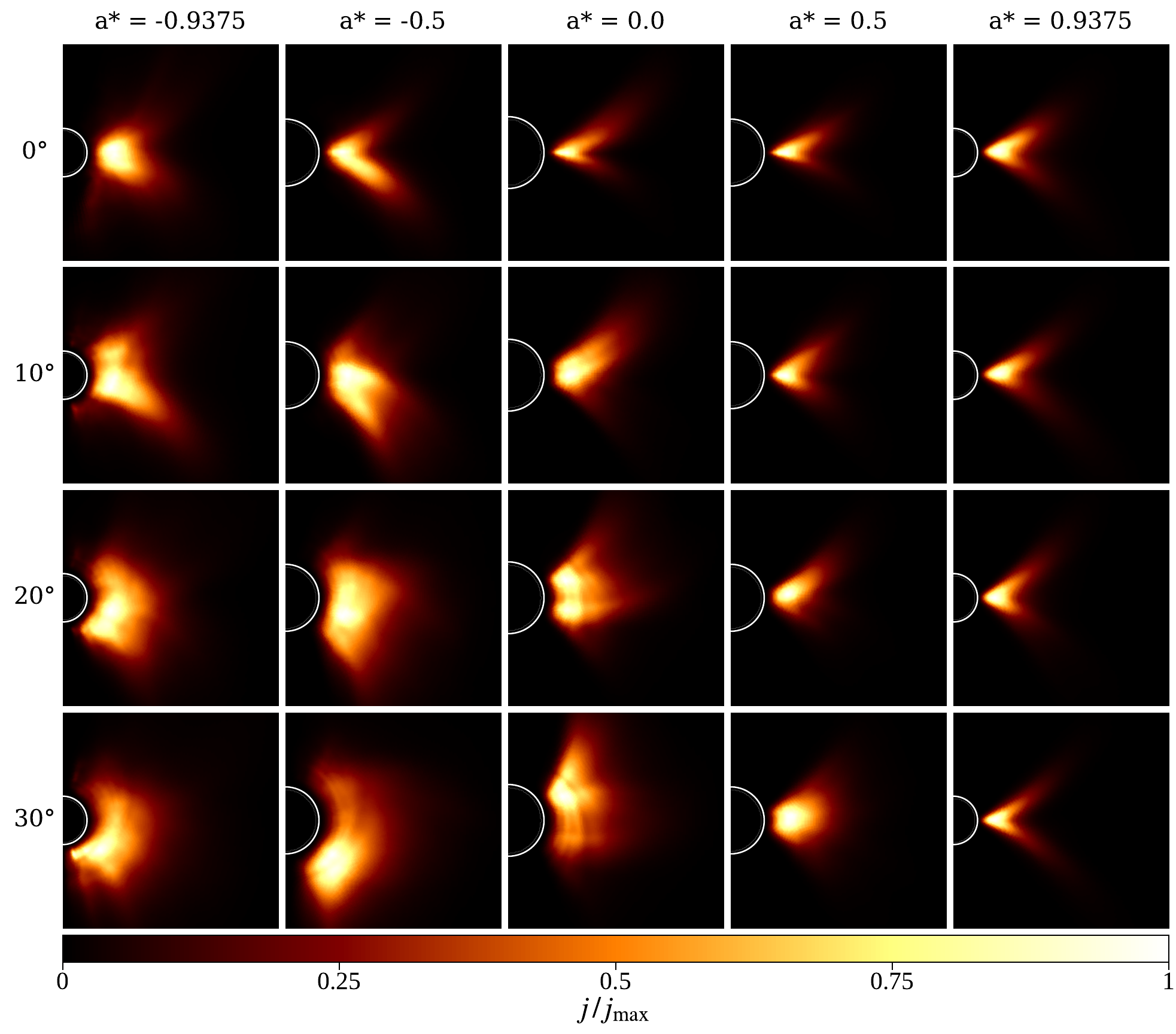}
    \caption{Binned location of the point of origin, or emissivities, for all photons that make up the individual images from \autoref{log_grid_230ghz_rh20_rl1_end02200}. The figure is identical to \autoref{emissivity_origin_phi_slice_grid_230ghz_rhigh20_rlow1}, except the photons are additionally summed over azimuth. 
    }
    \label{grid_emiss_230ghz_rh20_rl1_end02200}
\end{figure*}

To connect the image morphology to the emitting regions in the simulations, we also construct emissivity maps. The emissivity maps record the binned emission locations of the point of origin of all photons that make up each image. The maps are averaged over the same snapshots as the images and {\autoref{emissivity_origin_phi_slice_grid_230ghz_rhigh20_rlow1} is a slice at a particular azimuthal angle while \autoref{grid_emiss_230ghz_rh20_rl1_end02200} is averaged along azimuth}.

\subsection{Semi-Analytic Model}
\label{sec:semi_analytic_model}

\begin{deluxetable*}{llcc}
\tablecaption{Summary of {the 13} fitted semi-analytic model parameters and search ranges used in the image domain fitting procedure.  A more detailed overview of the dual cone parameters can be found in \citep{2024ApJ...974..143C}.  
\label{tab:krang_parameter_ranges}}
\tablewidth{0pt}
\tabletypesize{\normalsize}
\tablehead{
\colhead{Parameter} &
\colhead{Description} &
\colhead{Units} &
\colhead{Range}
}
\startdata
$\theta_{\rm g}$ & Mass-to-distance ratio & $\mu$as & $[1.0,6.0]$ \\
$a^*$ & Black hole spin & $\cdots$ & $[0,0.99]$ \\
$\theta_{\rm o}$ & Observer inclination & deg & $[0,179]$ \\
$\theta_{\rm s}$ & Cone opening angle & deg & $[40,90]$ \\
$r_{\rm peak}$ & Peak emission radius & $GM/c^2$ & $[1,10^3]$ \\
$p_1$ & Inner emissivity exponent & $\cdots$ & $[0,5]$ \\
$p_2$ & Outer emissivity exponent & $\cdots$ & $[0,5]$ \\
$\chi$ & Fluid velocity azimuthal angle in ZAMO frame & rad & $[-\pi,\pi]$ \\
$\iota$ & Magnetic field orthogonality angle in fluid frame & rad & $[0,\pi/2]$ \\
$\beta_{\rm v}$ & Fluid speed in ZAMO frame & $c$ & $[0.01,0.9]$ \\
$\Sigma$ & Spectral index & $\cdots$ & $[-1,5]$ \\
$\eta$ & Magnetic field tangential angle in fluid frame & rad & $[-\pi,\pi]$ \\
p.a. & Projected spin axis position angle & rad & $[-\pi,\pi]$ \\
\enddata
\end{deluxetable*}

A host of semi-analytic black hole image models assume that the emitting region is axisymmetric and aligned with the black hole spin axis. For example, \citet{2019PhRvD.100b4018G} studied the direct, lensed, and photon ring contributions to black hole images using optically thin disks with axisymmetric emission profiles. Similarly, the EHT analysis of the Sgr A* employed axisymmetric emission models in Kerr and non-Kerr spacetimes to relate the observed emission ring diameter to the underlying black hole shadow \citep{2022ApJ...930L..17E}. These models provide computationally efficient and interpretable descriptions of horizon scale emission, but their symmetry assumptions do not allow the emitting flow to be independently tilted or warped relative to the black hole spin.

We choose to use the dual cone emission model of \citet{2024ApJ...974..143C}, which leverages the ray tracing code \texttt{Krang} \citep{2024JOSS....9.7273C}, to fit the time averaged GRMHD images. In this model, emission arises from axisymmetric plasma embedded in the Kerr spacetime and constrained to flow along two oppositely oriented cones aligned with the black hole spin axis. The dual cone model is designed to reproduce horizon scale images from optically thin, low luminosity accretion flows around black holes. Here, its aligned geometry provides a controlled comparison which we can test whether the image signatures of disk tilt can be absorbed by changes in other model parameters or instead require explicitly {non-axisymmetric} degrees of freedom.

The spectral intensity $I_\nu$ emitted from each cone is written as the product of a double power law radial envelope and a polarized synchrotron intensity term, 
\begin{align}
    I_\nu(r)
        \propto\mathcal J(r, R, p_1, p_2 )\,P_\nu(\mathbf{B}, \mathbf{p}).\label{eqn:emissivity}
\end{align}
Here,
\begin{align}
    \mathcal J(r, R, p_1, p_2)
        &=\frac{(r/R)^{p_1}}{1+(r/R)^{p_1+p_2}}\label{eqn:double_power_law}
\end{align}
is the dimensionless emission profile, where $r$ is the emission radius, $p_1$ and $p_2$ are power law indices, and $R$ is a characteristic emission radius. The synchrotron intensity is given by
\begin{align}
\label{eqn:polarization}
    P_\nu
        &\propto \delta^{3+\Sigma}l_p 
        \lvert \mathbf B\rvert^{1+\Sigma}\sin^{1+\Sigma}\zeta,
\end{align}
where $\delta$ is the {Doppler} factor of the fluid frame as observed in the lab frame, $\Sigma$ is the spectral index, $l_p$ is the path length of the photon through the emitting surface, $\mathbf{B}$ is the spatial components of $b^\mu$, and $\zeta$ is the crossing angle between the photon momentum and magnetic field vector in the fluid frame. The crossing angle is defined by
\begin{align}
    \sin\zeta
        &=\frac{\lvert \mathbf{p}\times \mathbf{B}\rvert}{\lvert\mathbf p\rvert\lvert \mathbf B\rvert}\label{eqn:kcrossb},
\end{align}
where $\mathbf{p}$ is the photon three momentum. The fluid is constrained to flow along the conical surfaces, with its direction parameterized by the angle $\chi$ relative to the radial direction in the zero angular momentum observer (ZAMO) frame. Notice that the dual cone model only fits positive spins in \autoref{tab:krang_parameter_ranges}. This is a different convention from our GRMHD simulations, where the semi-analytic model encodes retrograde flows by the fluid velocity direction $\chi$, not the sign of the black hole spin. 
The magnetic field orientation is parameterized by the azimuthal and polar angles $\eta$ and $\iota$ in the fluid frame.
The dual cone model has previously shown that it can reproduce a variety of time averaged aligned GRMHD images and recover the black hole parameters in that setting \citep{2024ApJ...974..143C}. 
We use the dual cone model as an aligned comparison to test whether tilt introduces image structure that can be absorbed by changes in aligned model parameters or whether it produces image morphology that is not captured by the model.

We fit all 20 of the time averaged GRMHD images with the {dual cone model} within the {parameter} ranges defined in \autoref{tab:krang_parameter_ranges}. 
We have fixed the image center of the model to be aligned with the image center of the ray traced GRMHD, and normalized all images to have unit flux when fitting.
We perform a global optimization with  an adaptive differential evolution algorithm \citep{Choi2013AdaptiveCauchy} implemented in \texttt{blackboxoptim.jl} \citep{Feldt_BlackBoxOptim_2026}, using one minus the normalized cross correlation between the images of the dual cone model and the target GRMHD as our objective function. 
As the optimization is partly stochastic, we perform two independent runs of 100,000 iterations for each image and retain the fit with the highest normalized cross correlation. 
Since the model assumes an aligned, axisymmetric emission geometry, the differences in the best fit parameters and the simulation parameters quantify the degree to which tilt can bias aligned model interpretations.

\section{Results And Discussion}

\label{sec:results}

The image suite indicates that disk tilt impacts horizon scale images most clearly when the inner MAD flow remains misaligned. If magnetic torques are capable of driving the inner flow into alignment, then the resulting images remain close to those of aligned disk, even if the outer disk remains tilted. If the inner flow instead remains warped or non axisymmetric, then the tilt can imprint directly on the inner shadow and on the image morphology.

The weak tilt dependence is most apparent in the high spin prograde models shown in \autoref{log_grid_230ghz_rh20_rl1_end02200}. The $a^*=0.9375$ simulations show little dependence on the initial disk tilt. Across the full range of tilts, the time averaged images and the emissivity maps remain nearly identical, indicating that the emitting region close to the black hole is completely aligned. The fluid density structure shows the same alignment trend. The semi-analytic fits show the same behavior, as the images remain close to the aligned disk morphologies, the best fit black hole parameters are similar across disk tilt {and close to their true, simulated values}.

\begin{figure*}
    \centering
    \includegraphics[width=\textwidth]{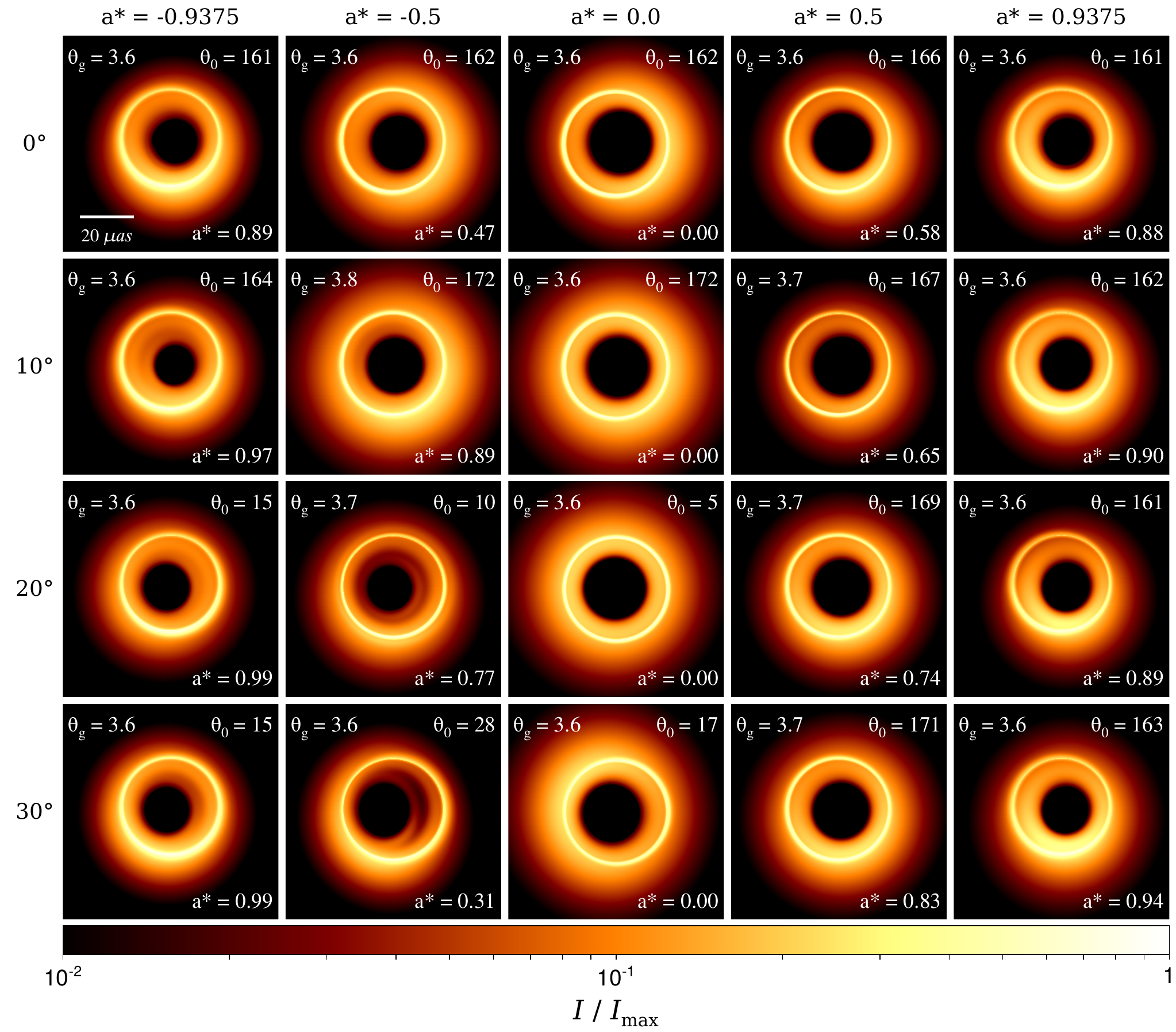}
    \caption{Best fit dual cone models to the \autoref{log_grid_230ghz_rh20_rl1_end02200} images, at 230 GHz with $R_{\rm high} = 20$ and $R_{\rm low} = 1$. Printed in the corners of each image are selected best fit parameters, where $\theta_{\rm g}$ is the mass to distance ratio in $\mu$as, $\theta_{\rm o}$ is the observer inclination in degrees, and $a^*$ is the black hole spin. As discussed in \autoref{sec:semi_analytic_model}, \texttt{Krang} fits only the absolute value of spin, as retrograde configurations are instead represented by changes in the fluid velocity direction. 
    }
    \label{bestfit_grid_230ghz_rhigh20_rlow1}
\end{figure*}

The intermediate spin prograde models show a similar but less complete version of the high spin alignment behavior. For $a^*=0.5$, the image morphology only depends weakly on disk tilt, suggesting that the inner accretion flow remains substantially aligned. The fluid density structure also indicates substantial inner flow alignment. However, the emissivity maps shown in \autoref{grid_emiss_230ghz_rh20_rl1_end02200} become slightly more extended in the most tilted cases. The extended emissivity structure indicates that a small residual non-axisymmetric structure survives in the emitting region. This is apparent in \autoref{emissivity_origin_phi_slice_grid_230ghz_rhigh20_rlow1}, where the emission remains slightly above the midplane. The semi-analytic fits in \autoref{bestfit_grid_230ghz_rhigh20_rlow1} reflect the same trend, where the inferred spin becomes increasingly biased at larger tilts, with larger tilts generally favoring higher fitted spins. 
Thus, even visually subtle tilt-induced image features can bias spin estimates.

The zero spin models provide a useful limiting case. With no black hole spin, there is no preferred alignment of the disk. For non-spinning black holes, changing the disk tilt is {trivially} equivalent to changing the viewing geometry  {because the spacetime is spherically symmetric}. \autoref{emissivity_origin_phi_slice_grid_230ghz_rhigh20_rlow1} supports this as the emissivity structure becomes simply tilted by the initial tilt angle. The resulting displacement of the central shadow is completely degenerate with change in observer inclination. The semi-analytic fits recover this geometric degeneracy and the black hole parameters are strongly recovered, while the fitted viewing inclination offsets the imposed disk tilt.

The retrograde models however exhibit very different behavior. As the tilt increases, the images show stronger changes in the size, shape, and position of the inner shadow. The inner shadow becomes less circular and in some cases {begins to shrink} and become warped. The emissivity maps also show strong non axisymmetric structure, with emission distributed over a broader region rather than concentrated in an azimuthally symmetric flow. The emissivity maps are consistent with the fluid averages shown in \autoref{rho_tilt30_spin-0.9375_2000_2020_avg}, which show a warped accretion structure that feeds the near horizon flow. In these cases, the tilted structure remains near the horizon and in the emitting region, imprinting directly on the images.

\begin{figure}
    \centering
    \includegraphics[width=3.3in]{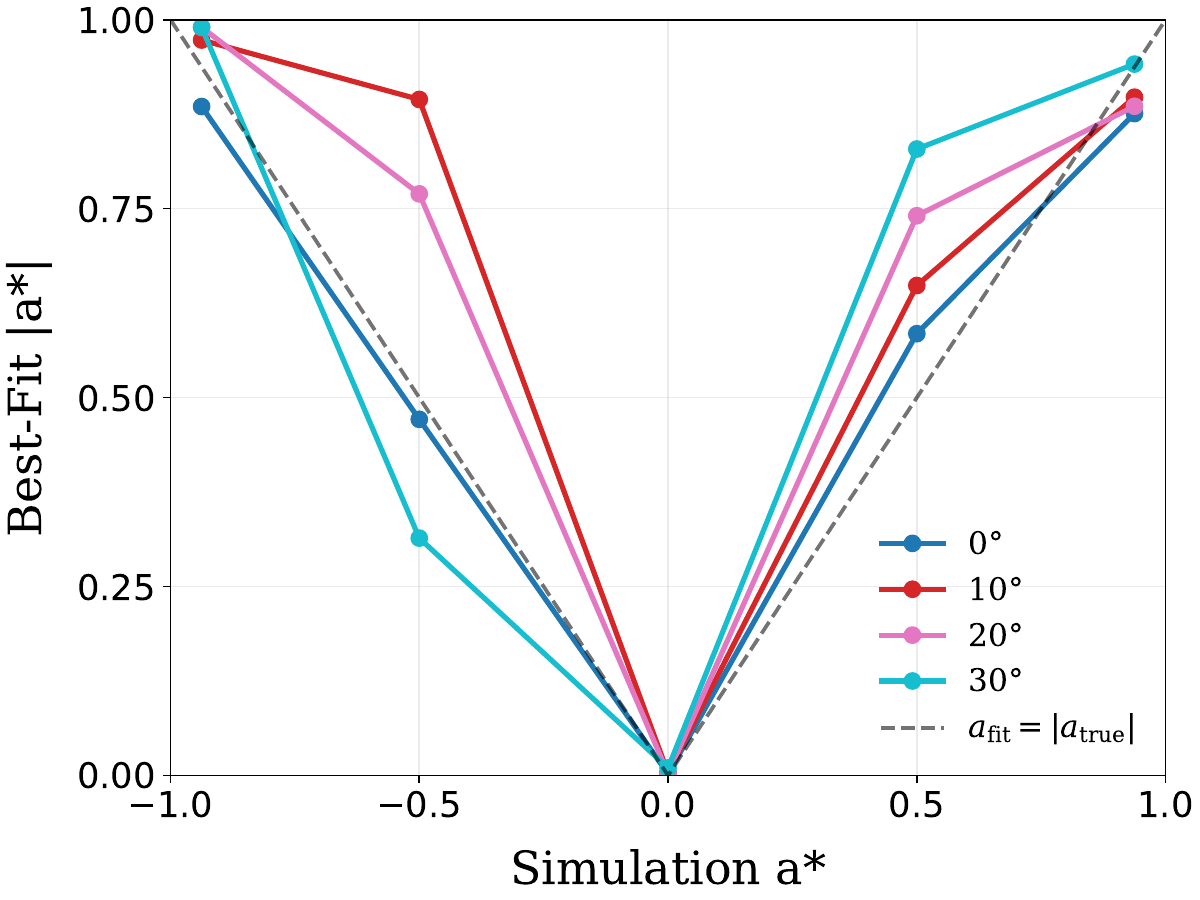}
    \caption{Best-fit model spins from \autoref{bestfit_grid_230ghz_rhigh20_rlow1} plotted against the true GRMHD spins for the images from \autoref{log_grid_230ghz_rh20_rl1_end02200} across all tilt angles. The dashed line indicates where the model would recover the true black hole spin values. The comparison uses the absolute value of the true black hole spin as \texttt{Krang} represents retrograde accretion by reversing the fluid velocity direction rather than by assigning negative spin.
    }
    \label{true_spin_vs_bestfit_spin}
\end{figure}

The semi-analytic fits further support the interpretation that the inner misalignment imprints on the image morphology. In the retrograde models, the fits generally degrade and the inferred black hole parameters, particularly the spin, become increasingly biased. This is shown in Figure \ref{true_spin_vs_bestfit_spin}, where values of the inferred spin can be biased by upwards of 50\% in specific cases. The biased parameters suggest that the aligned model attempts to compensate for missing tilted degrees of freedom by changing the black hole parameters. The aligned model cannot reproduce the warped, non-circular central brightness depression present in the retrograde tilted images, which likely contributes to the poorer fit quality.

\begin{figure}
    \centering
    \includegraphics[width=3.3in]{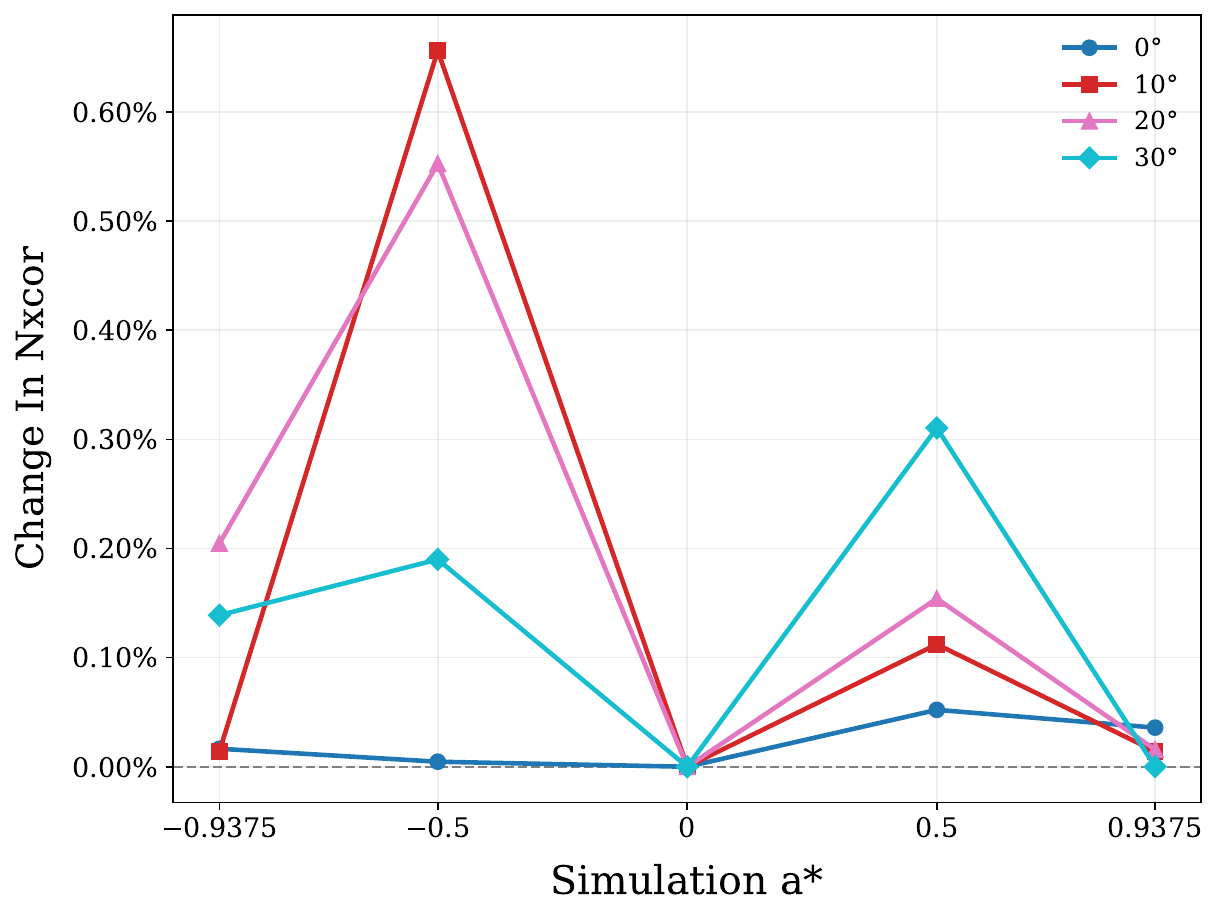}
    \caption{{Percentage decrease in the normalized cross correlation (Nxcorr) between best fit images in \autoref{bestfit_grid_230ghz_rhigh20_rlow1} and fits that fix the spin to its true value. The change in the Nxcorr plotted is signed; the values are only positive as the change can only get worse when we fix the spin.}
    }
    \label{nxcor_difference}
\end{figure}

{Because the fitted spin may be (partially or even fully) degenerate with other parameters in the semi-analytic model, we quantify the uncertainty of our fitted spins by refitting our images while fixing the spin to its true, simulation value, and calculating the fractional difference in the normalized cross correlation (Nxcorr) between the two fit images.  This provides a numerical measure to how much deviation there is in the image structure when the spin is held fixed, forcing the model to compensate for tilted disk image features with changes in disk and spacetime parameters.  The Nxcorr values with fixed spin are shown in \autoref{nxcor_difference}. In general, models with larger spin residuals also show larger fractional differences in Nxcorr when the spin is fixed to the simulated value, suggesting that the fitted model parameters are global minima. We find that the Nxcorr curvature near the minimum is ${\rm d}^2 {\rm Nxcorr} / {\rm d}a^2 \sim 0.1$, although the precise value varies considerably with spin and tilt. 
}

These results follow naturally from recent studies of tilted MAD disks. In particular, \citet{2026arXiv260116370G} found that high spin prograde tilted MADs align efficiently with the black hole spin, while retrograde tilted MADs remain persistently misaligned and develop warped accretion structures. Their results are consistent with the spin dependence shown in our simulations. The $a^*=0.9375$ prograde models produce nearly tilt independent images while the retrograde models retain warped inner structure that modifies the image morphology. Similar behavior is found by \citet{2025PhRvD.112f3013C}, who showed significant residual misalignment in the innermost $\sim 10$ $r_g$ of retrograde tilted MADs and strong alignment in their highest spinning prograde models. Although these studies do not focus on ray traced images, they identify the same underlying fluid behavior that controls the image morphology in our simulations.

Our results also demonstrate how fully MAD tilted disks differ from the SANE or semi-MAD flows considered in previous tilted disk imaging studies. Previous work showed that tilted disks can produce azimuthal dependent image structure, distort the central brightness depression, and that not including tilt in aligned models can bias parameters inferred \citep{2013MNRAS.432.2252D,2020ApJ...894...14W,2020MNRAS.499..362C}. In fully MAD flows, these signatures are not universal. In the highest spinning prograde MADs, magnetic torques suppress the observable imprint of large scale tilt by aligning the inner emitting accretion flow. In contrast, retrograde MADs retain an inner flow that can remain warped and non axisymmetric, allowing the tilt to imprint on the horizon scale image. Thus, the relevant condition for tilt induced image structure is not the presence of large scale disk misalignment, but whether the misalignment survives into the near horizon emitting region.

The dependence on residual inner misalignment has direct implications for interpreting black hole images. If the inner emitting region is completely aligned, then the aligned image models may provide a good description even when the large scale disk is tilted. However, if the inner flow remains warped, as shown in the retrograde tilted cases, aligned models can produce biased black hole parameters or fail to reproduce the detailed image morphology. The potential bias is especially important for spin inference, since spin is one of the most strongly affected parameters in the semi-analytic fits. Future semi-analytic models should therefore treat disk tilt as a source of horizon scale structure.

Several limitations of the present study should be kept in mind. The simulations are evolved to $11,000$ $t_g$, and the images are generated from the final $1,000$ $t_g$. Although the inner emitting region is expected to equilibrate on shorter timescales than the outer disk, a larger suite of long duration tilted MAD simulations would be useful for confirming the full spin-tilt trends. Second, we adopt a single observer azimuth prescription based on the time averaged magnetic pressure weighted centroid of the inner jet. The single azimuth prescription provides a controlled comparison across simulations, but tilted disks are intrinsically non-axisymmetric, and a full observer azimuth {and inclination survey} may reveal additional image structure. Additionally, different electron temperature prescriptions {and different observing frequencies} can emphasize different regions of the flow, and this may affect the strength of the observed tilt signatures.

\section{Conclusion}

\label{sec:conclusion}

In this work, we studied the image morphology of tilted magnetically arrested disks using a suite of GRMHD simulations spanning black hole spins $a^*=-0.9375$ to $0.9375$ and initial disk tilts from $0^\circ$ to $30^\circ$, with M87* parameters at a fixed $163^\circ/17^\circ$ inclination. We generated time averaged synthetic images and emissivity maps, and compared the 230 GHz images to an {axis-symmetric} semi-analytic model to test when disk tilt produces distinct image structure. Our main conclusions are as follows:

\textbf{High spin prograde MADs:} For $a^*=0.9375$, the image morphology is nearly independent of the initial disk tilt. The time averaged images and emissivity maps remain similar from $0^\circ$ to $30^\circ$. The weak tilt dependence is caused by magnetic torques from the jet that align the inner emitting {region} with the black hole spin axis. In this regime, aligned image models provide a good description {of the underlying geometry} when the large scale disk is tilted.

\textbf{Intermediate spin prograde MADs:} For $a^*=0.5$, the images remain broadly similar across tilt, but the most tilted cases show subtle changes in the emitting region. These residual differences are small visually, but they can still bias the parameters recovered by aligned semi-analytic fits, especially the black hole spin, {with spin bias of $\sim50$\% in the most tilted case}.

\textbf{Retrograde MADs:} For $a^*<0$, disk tilt produces much stronger changes in the image morphology. The inner accretion flow remains warped and non-axisymmetric, producing more extended emitting regions and distorted central brightness depressions. For retrograde tilted MADs, aligned semi-analytic models provide much poorer fits and can return substantially biased black hole parameters.

The imprint of disk tilt {on image structure} is not determined by the large scale disk orientation alone, but by whether misalignment survives into the near horizon emitting region. When the inner flow aligns, tilted MAD images can resemble aligned disks. When the inner flow remains warped {and non-axisymmetric}, tilt leaves distinct image signatures that aligned models cannot fully absorb. While we have focused our study on simulations viewed at $17^\circ$, as motivated by observations of M87*, the spin-dependent image morphology suggest that aligned image models are reliable for tilted MAD systems only when the emitting region has itself become aligned in the higher spin, prograde regime. For retrograde or otherwise persistently warped flows, disk tilt should be treated as an explicit degree of freedom in future semi-analytic modeling of horizon scale black hole images.

\section{ACKNOWLEDGMENTS}

We thank Joshua Stanway, Morgan MacLeod, Daichi Tsuna, Angelo Ricarte, Paul Tiede, and Charles Gammie for their helpful discussions and commentary. 
We acknowledge financial support from the National Science Foundation (AST-2307887). This publication is funded in part by the Gordon and Betty Moore Foundation, Grant GBMF12987 and Grant \#13526. It was also made possible through the support of a grant from the John Templeton Foundation (Grant \#63445).  The opinions expressed in this publication are those of the author(s) and do not necessarily reflect the views of these Foundations.
This research used resources of the Oak Ridge Leadership Computing Facility at the Oak Ridge National Laboratory, which is supported by the Office of Science of the U.S. Department of Energy under Contract No. DE-AC05-00OR22725. This research used resources of the Argonne Leadership Computing Facility, which is a DOE Office of Science User Facility supported under Contract DE-AC02-06CH11357.

\appendix
\twocolumngrid

\section{Disk Tilt Analysis}
\label{app:disk_tilt_analysis}

\begin{figure*}
    \centering
    \includegraphics[width=\textwidth]{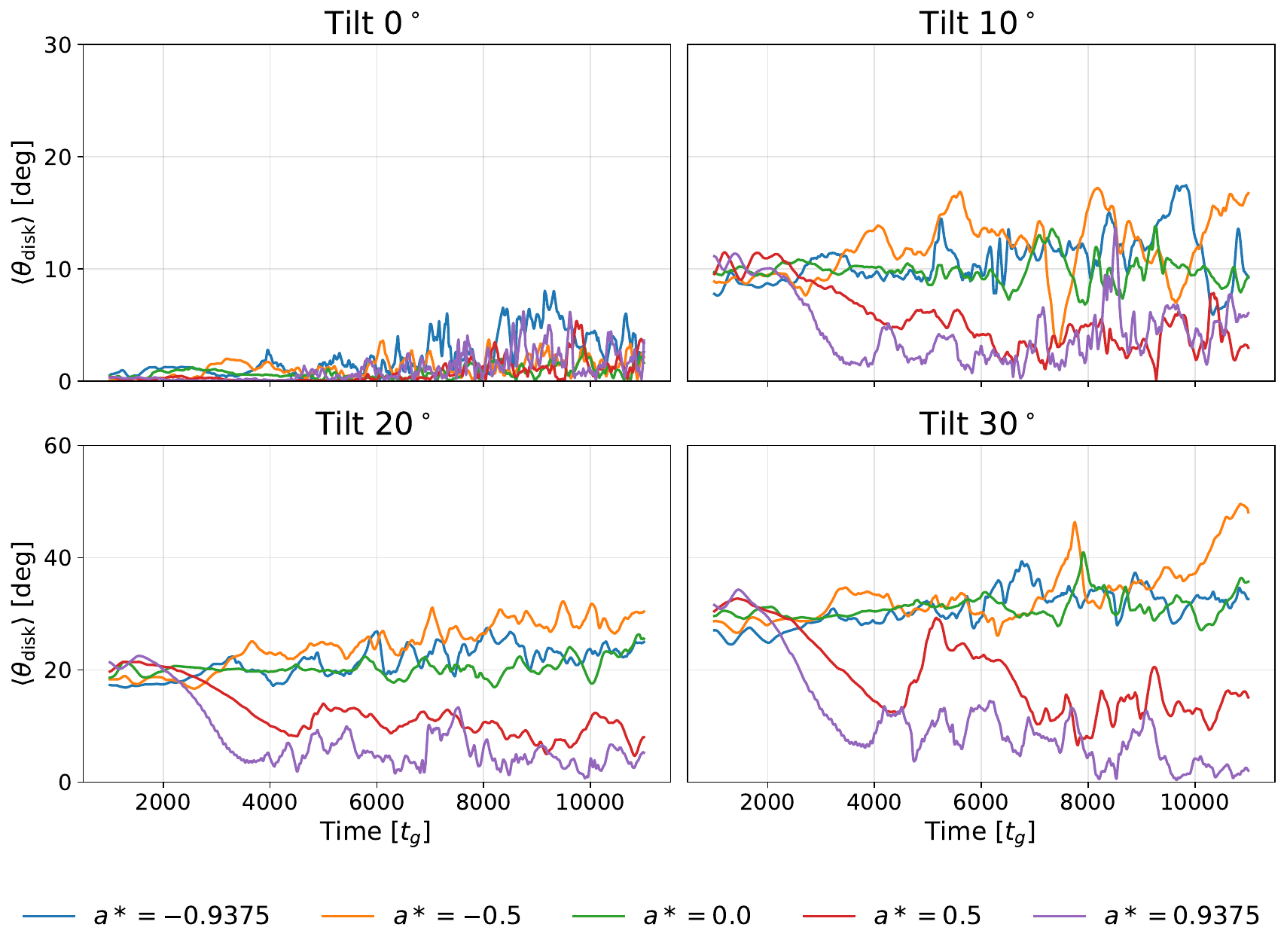}
    \caption{Time evolution of the inner disk tilt angle, $\langle \theta_{\rm disk} \rangle$, averaged over the radial range $5$ $r_g \leq r \leq 15$ $r_g$ for the full simulation suite. Each panel corresponds to a different initial disk tilt, and the colored curves show different black hole spins. }
    \label{DiskProfile_t00_t10_t20_t30_r5_to_15rg_theta_vs_tg_all_spins}
\end{figure*}

To quantify the time dependent orientation of the accretion flow, we compute the disk angular momentum vector as a function of radius and time. This diagnostic is used to measure whether the inner disk, and in particular the region that dominates the emission, has reached a steady state during the imaging interval and is no longer aligning. Since the black hole spin axis is aligned or anti-aligned with the coordinate polar axis in all simulations, depending on whether the black hole has positive or negative spin, the disk tilt angle is defined as the angle between the disk angular momentum vector and the simulation $+\hat{z}$ direction.

For each simulation snapshot, we first identify disk material using the same magnetization cut adopted in the ray tracing analysis, retaining cells with $\sigma < 1$. We then compute the disk angular momentum vector in concentric radial shells. Following closely the angular momentum prescription used in previous tilted disk studies \citep{2007ApJ...668..417F,2005ApJ...623..347F}, we define the shell four momentum as
\begin{equation}
    S^\sigma(r)
    =
    \int_{\Delta r}
    T^{\sigma 0}_{\rm fl}\,dV ,
    \label{eq:disk_shell_four_momentum}
\end{equation}
and the angular momentum tensor as
\begin{equation}
    L^{\mu\nu}(r)
    =
    \int_{\Delta r}
    \left(
        x^\mu T^{\nu 0}_{\rm fl}
        -
        x^\nu T^{\mu 0}_{\rm fl}
    \right)dV .
    \label{eq:disk_angmom_tensor}
\end{equation}
Here $\mu$, $\nu$, and $\sigma$ are spacetime indices, and the integrals are taken over disk cells in the radial shell $\Delta r$. The simulation volume element is
\begin{equation}
    dV = \sqrt{-g}\,dx^1 dx^2 dx^3 .
    \label{eq:disk_tilt_volume_element}
\end{equation}

We use the fluid part of the stress energy tensor,
\begin{equation}
    T^{\mu\nu}_{\rm fl}
    =
    \rho h u^\mu u^\nu
    +
    p g^{\mu\nu},
    \label{eq:fluid_stress_energy_enthalpy}
\end{equation}
where $\rho$ is the rest mass density, $h$ is the specific enthalpy, $p$ is the gas pressure, and $u^\mu$ is the fluid four velocity. 

The components entering Equations~\ref{eq:disk_shell_four_momentum} and \ref{eq:disk_angmom_tensor} are therefore
\begin{equation}
    T^{\mu 0}_{\rm fl}
    =
    \rho h u^\mu u^0
    +
    p g^{\mu 0}.
    \label{eq:fluid_tmu0}
\end{equation}

The disk angular momentum vector is obtained by projecting the angular momentum tensor with the shell four momentum,
\begin{equation}
    \left(J_{\rm disk}\right)_\rho
    =
    \frac{
        \epsilon_{\mu\nu\sigma\rho}
        L^{\mu\nu}S^\sigma
    }
    {
        2\left(-S^\alpha S_\alpha\right)^{1/2}
    } ,
    \label{eq:disk_angmom_projected}
\end{equation}
where $\epsilon_{\mu\nu\sigma\rho}$ is the Levi-Civita tensor. This expression gives the angular momentum vector associated with the disk material in each radial shell.

The black hole spin vector is taken to point along the simulation $+\hat{z}$ direction. Therefore the local disk tilt angle is computed by projecting $\mathbf{J}_{\rm disk}$ onto this axis:
\begin{equation}
    \theta_{\rm disk}(r,t)
    =
    \cos^{-1}
    \left[
        \frac{J_z(r,t)}
        {|\mathbf{J}(r,t)|}
    \right].
    \label{eq:disk_tilt_angle}
\end{equation}
With this convention, an aligned disk has $\theta_{\rm disk}=0^\circ$, while a disk retaining its initial tilt has $\theta_{\rm disk}$ close to the imposed tilt angle. This is the same tilt convention used throughout the paper.

To connect this diagnostic to the image morphology, we focus on the radial range 5 $r_g \leq r \leq$ 15 $r_g$. For each snapshot, we average the disk angular momentum vector over this radial interval and obtain a representative inner disk tilt angle, $\left\langle
        \theta_{\rm disk}(t)
    \right\rangle$.  
The resulting time series provides a direct measure of whether the emitting region is done aligning and has reached a quasi steady state during our imaging interval.

Figure~\ref{DiskProfile_t00_t10_t20_t30_r5_to_15rg_theta_vs_tg_all_spins} shows the evolution of this inner disk tilt diagnostic across the simulation suite. As shown in Figure~\ref{DiskProfile_t00_t10_t20_t30_r5_to_15rg_theta_vs_tg_all_spins}, the inner disk tilt angles generally exhibit little long term alignment or evolution during the final $1{,}000\, t_g$, corresponding to the imaging interval. This indicates that the orientation of the inner accretion flow has reached a broadly quasi-steady state by the time the synthetic images are generated.

The tilt angles nevertheless show short timescale variability. Some of these fluctuations are associated with magnetic flux eruption events, which can temporarily disrupt or partially tear the inner disk. These transient changes in the disk structure can produce some variability seen in the measured tilt angle without indicating continued long term alignment.

An additional caveat applies to the $a^*=-0.5$, tilt $30^\circ$ simulation. In this model, the disk occasionally extends into the polar region, where interaction with the reflecting polar boundary can shear part of the flow. This contributes to the large tilt angle fluctuations of $\sim 15^\circ$ near $t\sim7{,}500$ $t_g$ and again after $t\sim10{,}000$ $t_g$. Because the latter fluctuation overlaps with the imaging interval, the results from this simulation should be interpreted with this limitation in mind. We do not observe comparable polar boundary interactions in the remainder of the simulation suite, where the disks generally remain sufficiently far from the coordinate poles {to avoid tearing and shearing}.

\bibliography{references}{}
\bibliographystyle{aasjournal}

\end{document}